\pdfoutput=1
\documentclass[aps,prd,twocolumn,groupedaddress,nofootinbib,longbibliography]{revtex4-1}

\usepackage{amsfonts,amsmath,amssymb}
\usepackage{nicefrac}
\usepackage{graphicx}
\usepackage{color}
\usepackage{soul} 
\usepackage{hyperref}
\usepackage{accents}
\usepackage[normalem]{ulem}

\def\beq{\begin{equation}}
\def\eeq{\end{equation}}
\def\bea{\begin{eqnarray}}
\def\eea{\end{eqnarray}}

\begin{document}


\title{Binary Black Hole in a Double Magnetic Monopole Field \\}

\author{Maria J.~Rodriguez}
\email[]{maria.rodriguez@aei.mpg.de}
\email[]{maria.rodriguez@usu.edu}
\affiliation{Utah State University, Department of Physics, 4415 Old Main Hill Road, UT 84322, USA }
\affiliation{Max Planck Institute for Gravitational Physics, Am M\"uhlenberg 1, Potsdam 14476, Germany}



\begin{abstract}
Ambient magnetic fields are thought to play a critical role in black hole jet formation. Furthermore, dual electromagnetic signals could be produced during the inspiral and merger of binary black hole systems. However, due to the absence of theoretical models, the physical status of binary black hole arrays with dual jets has remained unresolved. In this paper, we derive the exact solution for the electromagnetic field occurring when a static, axisymmetric binary black hole system is placed in the field of two magnetic or electric monopoles. As a by-product of this derivation, we also find the exact solution of the binary black hole configuration in a magnetic or electric dipole field. The presence of conical singularities in the static black hole binaries represent the gravitational attraction between the black holes that also drag the external two monopole field. We show that these off-balance configurations generate no energy outflows.
\end{abstract}


\maketitle

\pagestyle{plain}

\section{introduction}
Dazzling electromagnetic signals can be produced during the inspiral and merger of a binary supermassive black hole system as a natural outcome of galaxy mergers \cite{2015ApJ...806..147C,2041-8205-752-1-L15}. The highly collimated jets of energy originate from a localized central region of the galaxies - generally believed to be spinning black holes - and are thought to play a fundamental role in the formation and evolution of massive galaxies.

In addition to the first direct detection of gravitational waves \cite{MariononbehalfoftheLIGOScientificCollaboration:2017vqx}, the first observation of the collision and merger of a pair of stellar mass black holes was recently reported by the Laser Interferometer Gravitational-wave Observatory (LIGO) signaling a major scientific breakthrough \cite{TheLIGOScientific:2016pea}. The inspiral of binary black hole systems is therefore of utmost astrophysical importance for future observations in the gravitational wave band as well as in the various electromagnetic bands. 

Alongside these substantial developments in the observation of binary black hole signals, there have been dramatic advancements in our theoretical understanding of electromagnetic emissions from black holes \cite{Tchekhovskoy:2009ba,Palenzuela:2010nf,Shapiro:2017cny,2041-8205-749-2-L32,Komissarov:2005wj}. The basic picture of the electromagnetic (force-free) model for black holes was developed by Blandford and Znajek (BZ) \cite{Blandford:1977ds} and entails an external magnetic field that is { \it dragged} by a rotating black hole background inducing electric fields, allowing currents to flow.

Extending the success of the process described by BZ for a single spinning black hole, the picture that emerges in \cite{Palenzuela:2010nf} for a binary black hole array with jets suggests that the dynamical behavior of the system induces collimation of the electromagnetic field around each intervening black hole, distilling energy from them and ultimately merging into a single black hole with the standard BZ scenario. 

In spite of all the progress, a detailed understanding of the jets of binary black holes remains elusive. In part, due to the inability of detecting clean electromagnetic signals from the central region close to the black holes and to the lack of theoretical models dealing with the interaction of electromagnetic fields and possible emissions in binary black hole systems.

As one step towards a systematic theoretical modeling of the electromagnetic emission from an inspiralling black hole binary, we consider the more simplistic scenario in which the binary is static and embedded in an external two magnetic monopole field. This is a reasonable model since, for all practical astrophysical purposes, the gravitational back-reaction of the magnetic field can be neglected. In this approximation, the electromagnetic field may be treated as a ''test'' field on the unperturbed, asymptotically flat, neutral binary black hole solution.
Specifically, we focus on a static binary black hole configuration - a pair of black holes at a finite distance with arbitrary masses - and find exact solutions for the two magnetic(electric) monopoles external field. Our starting point is a class of solutions to Einstein-Maxwell theory \cite{Emparan:2001bb,Alekseev:2007re} that represent two charged black holes. Employing a linearized analysis on the fully back-reacted solutions we find the static neutral binary black hole solution embedded in an external two monopole magnetic field. In \cite{Gibbons:2013yq} the authors employed the linearized analysis (starting with the Kerr solution) to recover Wald's solution \cite{Wald:1974np}. The features of the binary black hole in a two magnetic monopole field provide a connection to several of the above mentioned issues. In particular, we will study these new exact electromagnetic fields to test the results of energy outflows in head-on collisions of non-spinning black holes \cite{Neilsen:2010ax,Palenzuela:2010xn}. The exact electromagnetic field solutions that we construct are very general (with arbitrary masses and charges) and may be relevant for initial data in numerical simulations of orbiting binary black holes.

\section{Setup}

As we previously discussed, we will focus in the neutral black hole solutions of $R_{\mu\nu}=0$ in electromagnetic field backgrounds $F=dA$, with gauge field $A$, that solve Maxwell's equations in vacuum
\beq\label{maxwell}
dF=0\,,\qquad d*F=0\,. 
\eeq
We will adopt Weyl's canonical coordinates $(t,\rho,z,\phi)$, and assume the spacetime line element takes the form 
\beq\label{ds2}
ds^2 =- f\,dt^2+f^{-1}\,[e^{2\gamma}(d\rho+dz)+\rho^2 d\phi^2], 
\eeq
The functions $f, \gamma$ only depend on $\rho$ and $z$.

We begin with a review of the classic Michel magnetic monopole solution \cite{michel1973mon}, in polar and canonical Weyl coordinates, which illustrates the basic form of the electromagnetic test field for a single black hole. The exact binary static black hole solution \cite{IsraelKhan}, that is the central focus of this paper, is then discussed. We present our results for the black hole binaries in different double monopole electromagnetic fields, while the last section includes our assessment of the new configurations and conclusions.
 
\section{Schwarzschild Black Hole in Monopole Field}
To illustrate the more complex cases of background electromagnetic fields in binary black holes, we first start inspecting the magnetic monopole solution in a Schwarzschild black hole background
\beq\label{Sch}
ds^2=-\frac{r-2m}{r} dt^2+\frac{r}{r-2m}dr^2+r^2(d\theta^2+\sin^2\theta d\phi^2)\,.
\eeq
In these coordinates, the gauge field
\beq\label{magnetic}
A_{\mu}=(0,0,0,-q\cos\theta)\,.
\eeq
is a solution of (\ref{maxwell}) in the presence of a static black hole (\ref{Sch}) (which includes flat space-time as a special case when $m=0$). The corresponding field strength $F=dA$ is the monopole field of charge $q$ and has a constant flux integral $4\pi q$ for any radius.

The electric monopole field $\tilde{F}\equiv d\tilde{A}=* F$ with
\beq\label{electric}
\tilde{A}_{\mu}=({-2 q}/{r},0,0,0)\,,
\eeq
is also a solution of vacuum Maxwell's equations in the presence of a static black hole (\ref{Sch}).
 
 \subsection{Schwarzschild solution in Weyl coordinates}

The static binary black hole solutions that are the subject of this work are better represented in Weyl coordinates. Hence, it will be useful to review the description of the single monopole field solutions previously discussed in these coordinates. We start by inspecting the static Schwarzschild black hole in the Weyl canonical coordinates. To this end we transform the $(r,\theta)$ coordinates into  $(\rho,z)$ through
\beq\label{coord}
\rho=\sqrt{r^2- 2m\, r} \sin\theta\,,\qquad z=(r-m) \cos\theta\,.
\eeq
Then space-time metric (\ref{Sch}) becomes (\ref{ds2}) with functions 
\beq\label{SchWeyl}
f=\frac{R_+ + r_+ - 2 m}{R_+ + r_+ +2m}\,,\,\,\,
e^{2\gamma}=\frac{(R_+ + r_+)^2- 4 m^2}{R_+ r_+}\,,
\eeq
where we have defined
\beq
R_+=\sqrt{\rho^2+(z+m)^2} \,,\qquad r_+=\sqrt{\rho^2+(z-m)^2}\,.
\eeq
In Weyl coordinates, the black holes event horizon correspond to {\it rods} along the $z$-axis with $\rho=0$. For the Schwarzschild black hole the rod that represents the event horizon, has a of length equal to $2m$ and is located at $\{\rho=0, -m\le z\le m\}$.

\begin{figure}
  \includegraphics[width=4.2cm]{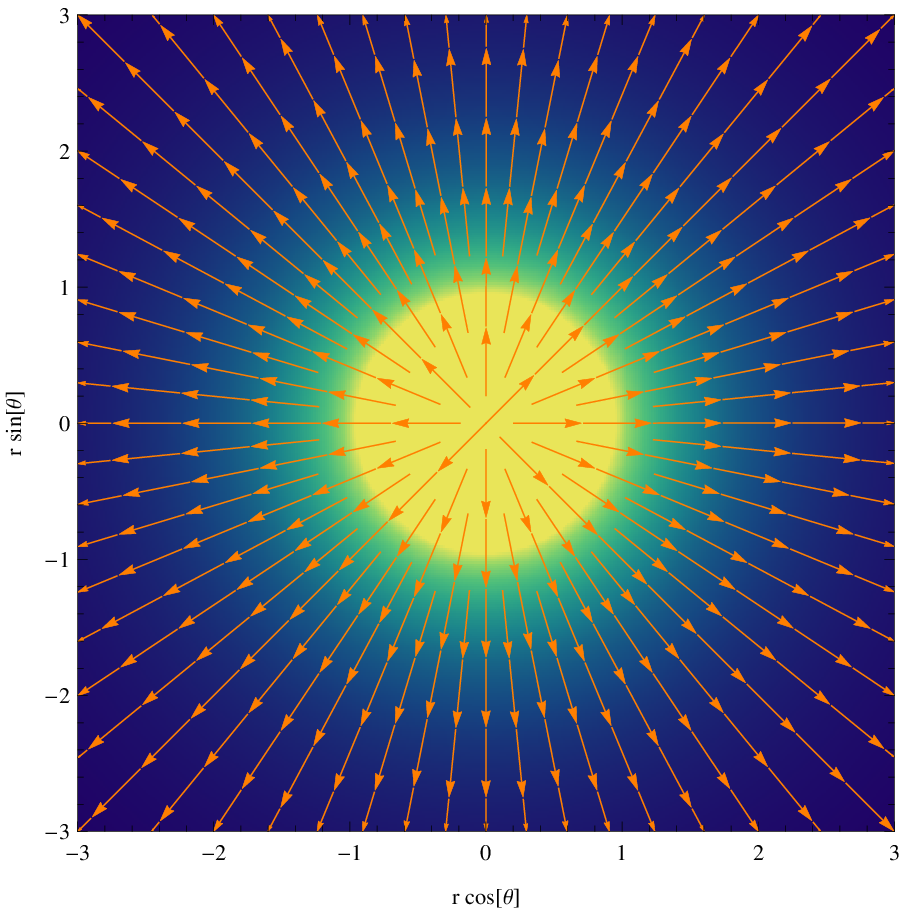}
   \includegraphics[width=4.2cm]{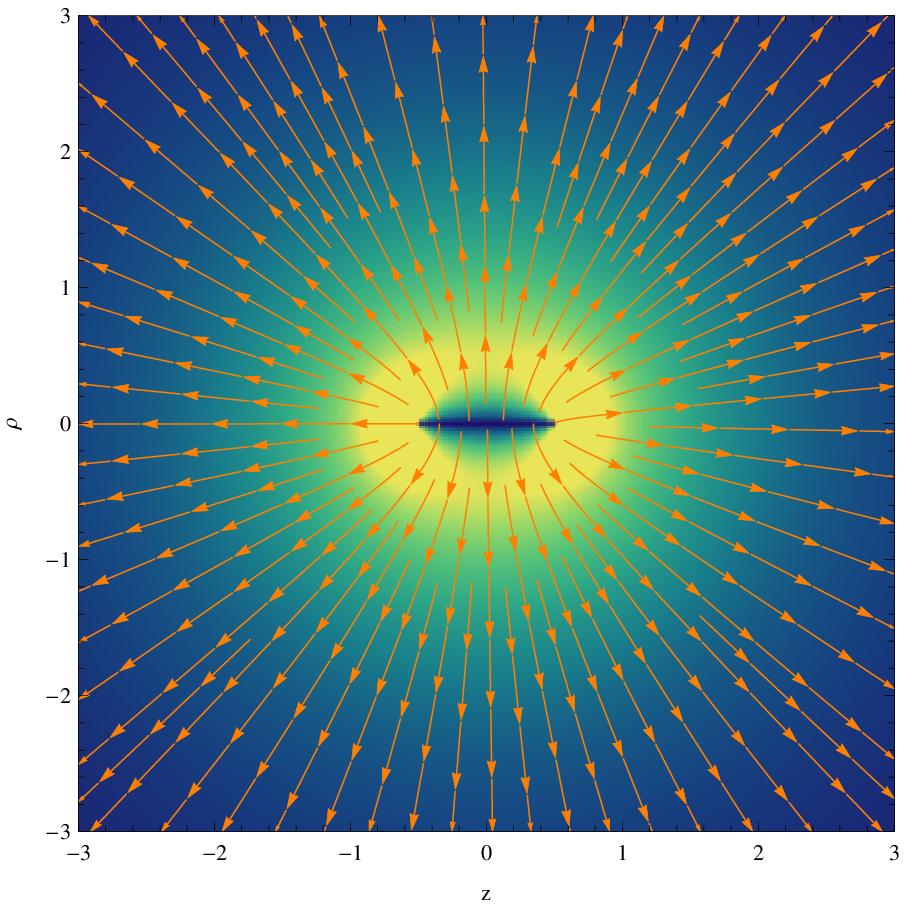}
  \caption{Electromagnetic structure of a radial magnetic monopole field ({\it in orange}) in a Schwarzschild black hole represented in cartesian {\it (left)} and Weyl {\it (right)} coordinates.}
  \label{fig:monopole}
\end{figure}

The magnetic monopole (\ref{magnetic}) in these coordinates yields
\beq\label{magneticW}
A_{\mu}=\left(0,0,0,  \frac{q(R_+ -r_+)}{2m}\right)\,,
\eeq
while the electric monopole field (\ref{electric}) becomes
\beq\label{electricW}
\tilde{A}_{\mu}=\left(\frac{-2q}{R_+ +r_+ +2m},0,0,0\right)\,.
\eeq
The electromagnetic structure of the background magnetic monopole field is depicted in Fig. \ref{fig:monopole}. Note that in Weyl coordinates the structure of the event horizon and the field lines slightly change.
 %

\section{Two Schwarzschild black holes Solution}

Now let us consider the Israel-Khan solution \cite{IsraelKhan}, which is an exact solution of Einstein's equation in vacuum describing a set of two static neutral black holes. This binary black hole system contains two black holes of arbitrary masses $m_1$ and $m_2$ and the solution is of the form (\ref{ds2}) with functions
\bea\label{binary}
f=\dfrac{\mathcal{D}-\mathcal{G}} {\mathcal{D}+\mathcal{G}}\,,\,\, \,e^{2\gamma}=\dfrac{f_0(\mathcal{D}^2-\mathcal{G}^2)}
{(x_1^2-m_1^2 y_1^2)
(x_2^2-m_2^2 y_2^2)}\,,
\eea
where $\mathcal{D}, \mathcal{G}$ are the polynomial functions
 \begin{equation}\label{DG}
\begin{array}{l}
\mathcal{D}=x_1 x_2+\delta\bigl[ x_1^2+x_2^2-(m_1 y_1-m_2 y_2)^2 \bigr]\\[1ex]
\mathcal{G}=m_1 x_2+m_2 x_1\\
\phantom{\mathcal{G}=}+2\delta\left[m_1 x_1+m_2 x_2-m_1 y_1\ell+m_2y_2\ell\right]\,,
\end{array}
\end{equation}
with parameters $\ell=z_2-z_1$ that characterizes the $z$-distance separating these sources (when  $0<z_1<z_2$), $ \delta= m_1 m_2/(\ell^2-m_1^2-m_2^2)$ the conical angle on the $z$-axis and, $f_0=(1+2 \delta)^{-2}$
guarantees the asymptotic flatness of the solution. Here $(x_i,y_i)$ with $i=1,2$ are the bi-polar coordinates centered on the symmetry axis $\rho=0$ at $z=z_1$ and $z=z_2$ defined by the expressions
\bea
x_i=\frac{R_i+r_i}{2}\,,\qquad  y_i=\frac{R_i-r_i}{2 m_i}\,,
\eea
where
\bea
R_i&=&\sqrt{\rho^2+(z+(m_i-z_i))^2}\,, \\ r_i&=&\sqrt{\rho^2+(z-(m_i+z_i))^2}\,.
\eea
The latter equations give an explicit expression of the binary static black hole solution. It is fully determined in terms of three free real parameters: besides $m_1,m_2$ that characterize the individual masses of the black holes $\ell$ is the separation between the black holes. We are considering the case $\ell>m_1+m_2$, which corresponds to two non-overlapping rods. The event horizons or rods are located at $\{\rho = 0, z_i - m_i \le z \le z_i +m_i\}$ where $z_i$ is the $i$-th rod mid point. The black holes are of course expected to attract each other. Hence, for any choice of these parameters the solution has conical singularities which indicate the presence of pressure, a {\it strut}, along the symmetry axis between the sources. Note that when $\ell=m_1+m_2$ , the rods overlap and the solution for an individual Schwarzschild black hole of mass $m_1+m_2$ can be retrieved. The Israel-Khan solution \cite{IsraelKhan} of two identical static (neutral) black holes as found in \cite{Emparan:2001bb} is recovered considering $m_1=m_2=m$ positioned at $-z_1=z_2=k$. 
Finally, it is worth noticing that the irreducible mass in e.g. the latter case is $M_{irr}=(\mathcal{A}/16\pi)^{1/2}$ where $\mathcal{A}=16\pi m^2 (1+m/k)$ is the event horizon area of a single black hole. Thus, while from a static neutral black hole energy cannot be extracted, the maximum extractable energy for the binary black hole configuration that we are studying yields $M-M_{irr}=(2-\sqrt{1+m/k})m$.

\section{Static binary black hole in Two Electric Monopole Field}
The most general five parameter solution that describes the superposition of two Reissner-Nordstr\"om sources with arbitrary parameters along the axis was presented in \cite{Alekseev:2007re}. This is a fully back-reacted solution to Einstein-Maxwell's equations that represents two charged black holes. We are interested in finding the two monopole electric field of the solution for two neutral black holes (without any back-reaction). Henceforth, employing a linearized analysis starting with a the two charged black hole solution, we are able to find the two electric monopole field  solution of (\ref{maxwell}). In this approximation the metric becomes precisely the neutral static binary black hole metric (\ref{ds2}) with functions (\ref{binary}), and 
\bea\label{gaugefield}
\tilde{F}=d\tilde{A}\,,\qquad \tilde{A}_{\mu}=\left(\dfrac{\mathcal{F}} {\mathcal{D}+\mathcal{G}},0,0,0\right)\,,
\eea
that is the gauge potential with $\mathcal{D},\mathcal{G}$ defined in (\ref{DG}),
\bea\label{F}
\mathcal{F}&=&2\delta[q_1 x_1+q_2 x_2+y_1(m_2\gamma_F-q_1\ell)+y_2(m_1\gamma_F+q_2\ell)]\nonumber\\
&&+q_1 x_2+q_2 x_1+\gamma_F(m_1 y_1+m_2 y_2)
\eea
and $\gamma_F= (m_2 q_1-m_1 q_2)/\ell$. This field defines a two electric monopole field for a static binary black hole system, each one parametrized by $q_1$ or $q_2$. For $q_1,q_2>0$ or $q_1,q_2<0$ the configurations carry a two electric monopole field and in the case of opposite signs e.g. $q_1<0$ and $q_2>0$ these carry an electric dipole field.

With these solutions in hand, we will turn to the magnetic case, which is motivated by a possible manifestation of the BZ process during the collision of black holes binaries resulting from galaxy mergers.

\section{Static binary black hole in Two Magnetic Monopole Field}
\begin{figure}
  \includegraphics[width=4.2cm]{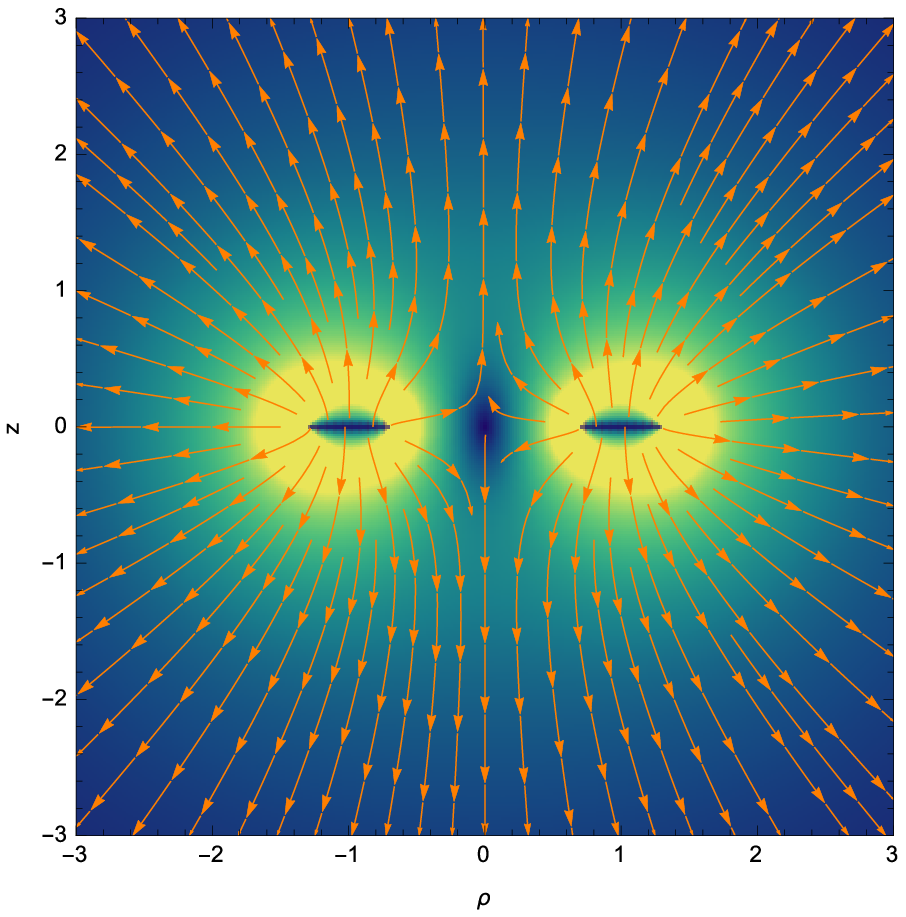}
   \includegraphics[width=4.2cm]{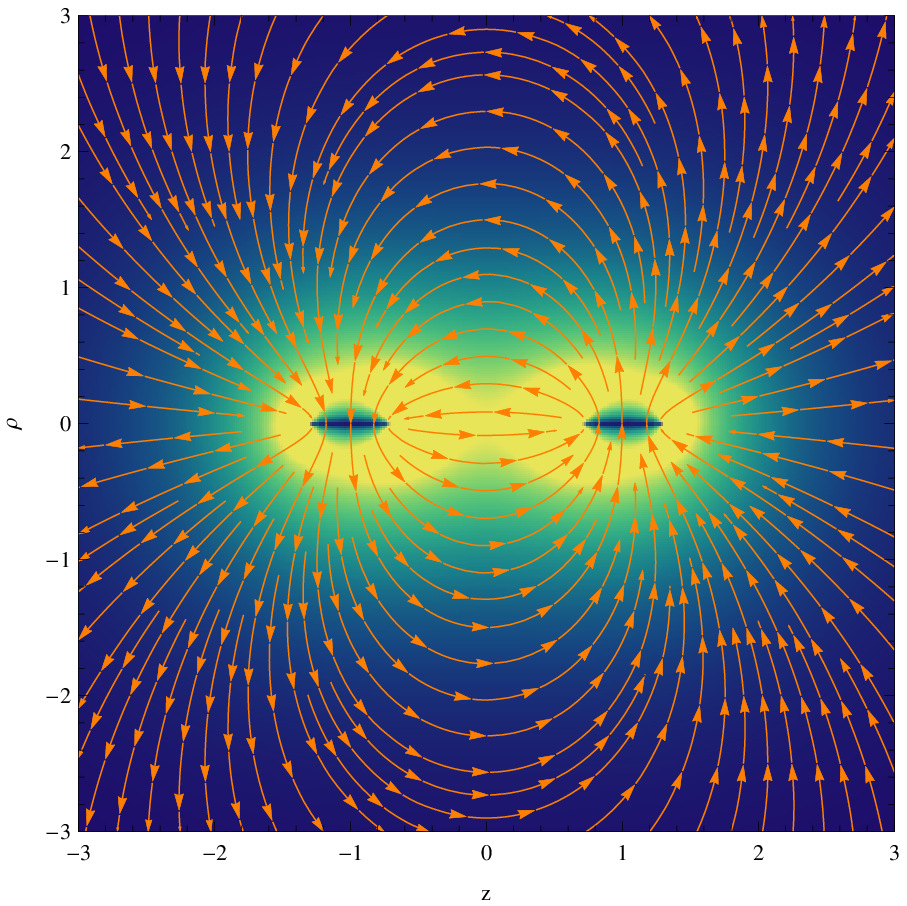}
  \caption{Electromagnetic structure of a two like magnetic monopole field ({\it left}) and a magnetic dipole field ({\it right}) in a (equal mass) binary black hole system.}
  \label{fig:monopole2}
\end{figure} 
To find the corresponding two magnetic monopole field we have to perform the duality transformation of the purely electric field (\ref{gaugefield}). The purely magnetic field
\bea\label{magneticfield}
F=-\frac{\rho}{f} d\phi \wedge (\partial_{z}\tilde{A}_t \, d\rho-\partial_{\rho}\tilde{A}_t \,dz)\,.
\eea
is a new exact solution of (\ref{maxwell}) when a static, axisymmetric binary black hole system (\ref{ds2}) with (\ref{binary}) is placed in this two magnetic monopoles field. 
As in the electric case, each magnetic field is governed by the free parameters $q_1,q_2$. Same charge sign configurations $q_1,q_2>0$ or $q_1,q_2<0$ carry a two magnetic monopole field and in the case of opposite signs e.g. $q_1<0$ and $q_2>0$ these correspond to dipole magnetic fields. Fig \ref{fig:monopole2} shows the electromagnetic structure of these fields for $q_1=\pm q_2$ in black hole binaries of the same size $m_1=m_2$.  As in \cite{Cazares:2007jn}, the construction of the magnetic potential $A_\phi$ can be carried out using Sibgatullin's method \cite{Sibgatullin:1991qr}. 
\begin{figure}
  \includegraphics[width=4.2cm]{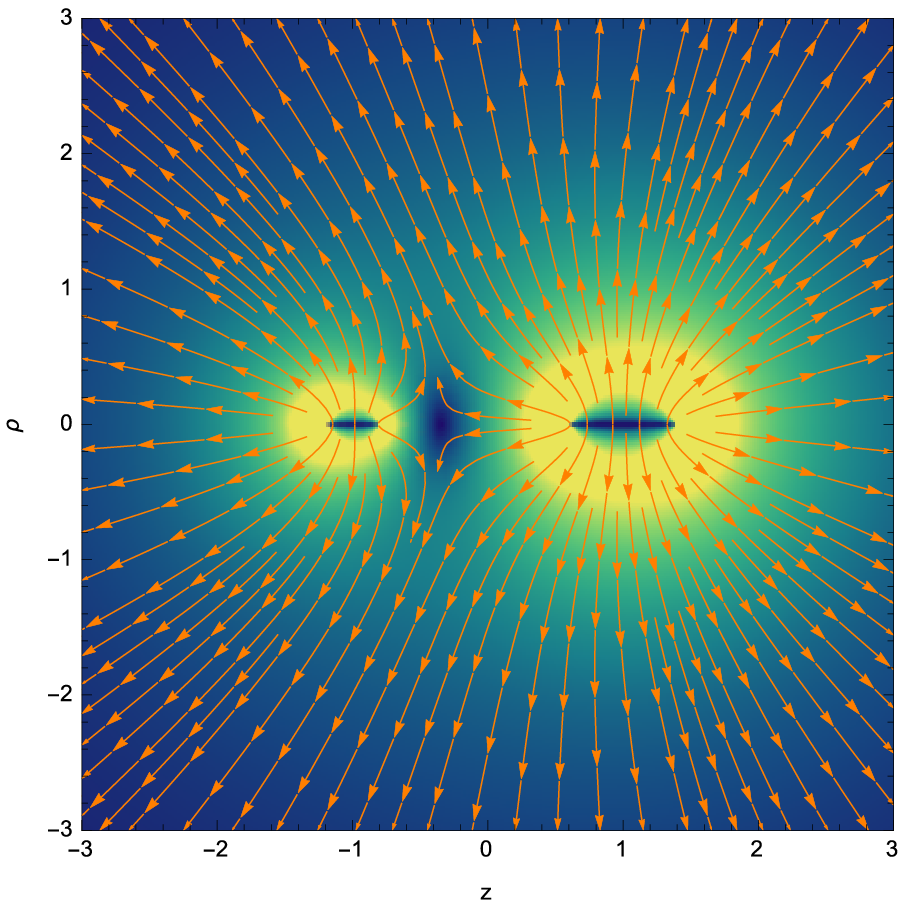}
      \includegraphics[width=4.2cm]{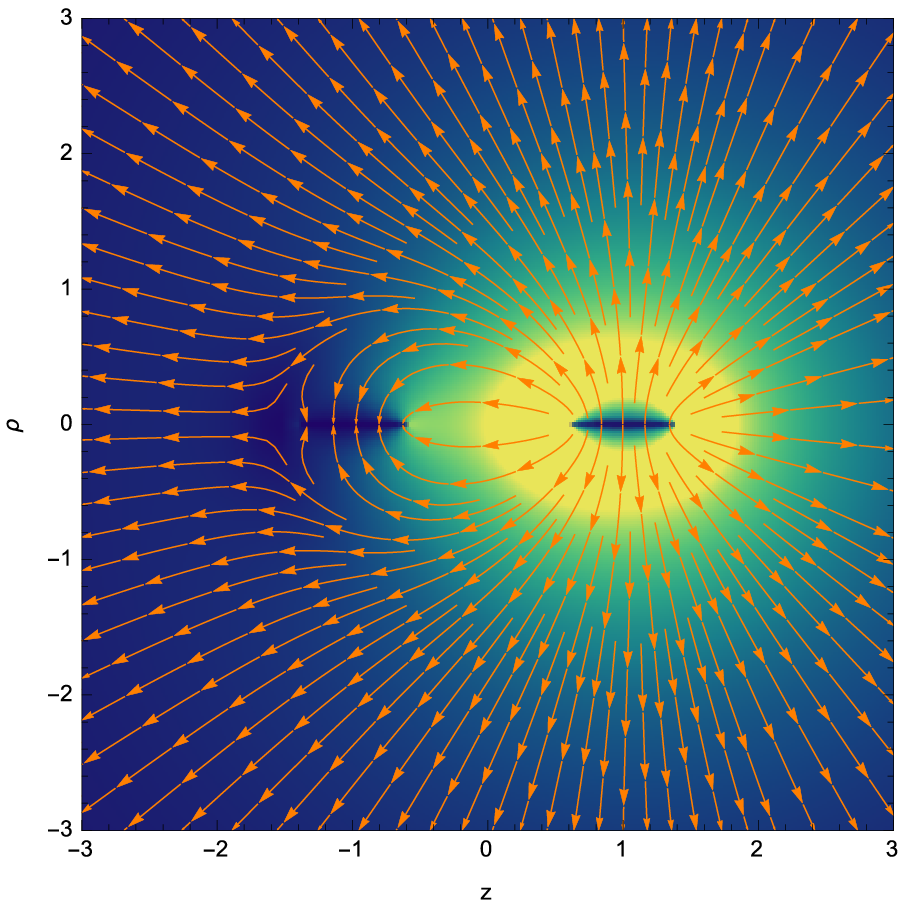}
  \caption{Electromagnetic structure of a two magnetic monople fields in a binary black hole system of unequal mass and charge ({\it left}) and one monopole field on (equal mass) binary black hole system ({\it right}).}
  \label{fig:monopole3}
\end{figure}
The new solutions that we find support two magnetic monopole fields of any magnitude for black holes of all sizes. 
As in the classical two magnetic monopole configurations (without black holes) the magnetic field lines radiate away(toward) such a positive(negative) magnetic monopole which can be thought of as an isolated magnetic north(south) pole. Another interesting situation is the limiting case in which one of the magnetic sources is zero e.g. $q_1=0$. This remains physically sensible and corresponds to two Schwarzschild black holes hovering freely in a one magnetic monopole field of charge $q_2$. The magnetic field patterns generated by both types of monopole are sketched in Fig. \ref{fig:monopole3}. Finally, one can recover the electromagnetic field of two point-like monopoles in Minkowski vacua when $m_1=m_2=0$.

The apparent singularities at the event horizons are simply discontinuities in physical coordinates. For the solutions  (\ref{magneticfield}) we find that $F^2$ is smooth on and outside the event horizons. In particular, for black hole binaries of the same mass ($m_1=m_2=m$) and equal external field ($q_1=q_2=q$) located at $z_1=-z_2=-k$ we find that $F^2|_{\rho=0,z=z_i\pm m_i}=k^2 q^2/8m^4(k+m)^2$.
 
As we mentioned earlier, the presence of conical singularities in the static (neutral) binary black hole solution that we are studying, indicates the presence of pressure along the $z$-axis.
It was shown in \cite{Costa:2000kf,Emparan:2001bb} that this gravitational pressure can be identified with the interaction energy. In our case this is
\bea\label{interaction}
V_{int}&\equiv& -\frac{\delta}{4} \int dz\,N\sqrt{g_{zz}}\, |_{\rho=0}\nonumber\\ 
&=&- \delta (1-\delta)^{-1} (\ell-m_1-m_2)/ 4\,,
\eea
where $N = \sqrt{- g_{tt}}$ is the lapse function for the static metric under consideration. We regard the pressure as a function of the distance $\ell$ between the black holes.  From (\ref{interaction}), we can study the effect of an external force between the constituent black holes by taking it out of equilibrium considering $\ell=\ell^{equil}+\Delta\ell$. The conical singularity for the static black hole binaries cannot be removed and the static black hole system is always unstable $V_{int}'<0$. Note that when the black holes are far apart $\ell>>1$ the results agree with Newtonian gravity expectations. Interestingly enough, while the system is off-balance in this stationary configuration, the two monopole magnetic field moves accordingly with the neutral binary black hole configuration. Effectively this can be interpreted as the dragging of the electromagnetic field by the black holes. Consequently, as shown in the head-on collisions of static binary black holes in \cite{Neilsen:2010ax,Palenzuela:2010xn}, these electromagnetic fields may drive non-trivial outflows of energy from the binary black holes with conical singularities. To test this hypothesis, we construct the split two monopole field solution for the binary black holes in analogy with the original BZ paper \cite{Blandford:1977ds}  (simply reversing the sign of the monopole charge across the equatorial plane $\rho=0$, entailing in this way an equatorial current sheet which accounts for a crude model of a disc). Computing the electromagnetic stress-energy tensor $T^{EM}_{\mu\nu}=F^{\mu\alpha}F_{\alpha}^{\nu}-(1/4)\,g^{\mu\nu}F_{\alpha\beta}F^{\alpha\beta}$ associated with (\ref{magneticfield}) we can find the net energy flux leaving the system via $\mathcal{E}\equiv \int  \, T^{\rho\, t}_{EM}  \sqrt{-g} \,d\phi$.
In spite of the split construction of the two monopole field, we find that the net energy outflow vanishes. Therefore, the evolution of the off-balance static binary black hole (with conical singularities) in the two magnetic monopole field does not conform to the expectation of energy outflows in the $\rho$-direction. These findings seem to be in contradiction with \cite{Neilsen:2010ax,Palenzuela:2010xn}. The authors considered the head on collision of non-spinning black holes, each embedded in an external constant magnetic field such as Wald's \cite{Wald:1974np}, and found non trivial energy outflows in the transverse direction e. g.  black holes moving towards each other in the $z$-direction in Weyl coordinates and outflows in the $\rho$-direction. Nevertheless, since the time evolution of the system in the numerical simulations presumably happen through highly-dynamical geometries, it is very likely that these have little to do with the conically-deformed ones that we have studied.

Let us now make a final comment about the magnetic field (\ref{magneticfield}) with (\ref{binary}). On the $z$-axis along the black holes, the lines of force are perpendicular and also vary. One can interpret this in terms of the standard magnetic pressure $\mathcal{P}_z (z)=(F^2/8\pi)|_{\rho=0}$ (which applies since the electromagnetic field is degenerate $*F^{\mu\nu}F_{\mu\nu}$=0) and check that the conical singularity in the metric has no effect in the magnetic pressure. This can be quickly verified by noting that the pressures at the ends of the rods are equivalent $\mathcal{P}_z(z=z_i- m_i)=\mathcal{P}_z(z=z_i+ m_i)$ causing no effect on the balance of the system.  At the origin, the magnetic pressure vanishes $\mathcal{P}_z(z=0)=0$.


\section{Discussion}

In this paper, we have obtained explicit analytic solutions describing the electric/magnetic field of two monopole fields in a static binary black hole configuration. The geometry is conically-deformed, and captures the dragging of the electromagnetic fields but not the energy outflows of the dual jets found in \cite{Neilsen:2010ax,Palenzuela:2010xn}. A closer approximation to real magnetospheres with non-trivial energy fluxes could be constructed by not simply splitting the monopole field, as we also did in this work, but also considering spinning black holes. We have not attempted here this generalization but hope to return to this subject in the future.

Our present construction is a first attempt for a theoretical model for dual jet emissions from binary black hole system in a two monopole field. Emissions along dual jets are expected during the merger of black holes binaries resulting from galaxy mergers and could be observable at large distances \cite{Cazares:2007jn}. Numerical studies seem to also indicate this possibility, although so far they have only considered black hole binaries with constant uniform test fields. It would be interesting to employ the newly found solutions as initial data in numerical simulations of orbiting binary black holes.

\begin{acknowledgments}
\end{acknowledgments}
I would like to thank the participants of the {\it ''Joint Columbia-USU Strings and Black Holes Workshop''}, 1-3 May 2017, in particular, Eric Hirschmann, Robert Penna and Bob Wald for inspiring discussions. I am grateful to Luis Lehner and Oscar Varela for various insightful comments. This work was supported by the Max Planck Gesellschaft through the {\it Gravitation and Black Hole Theory} Independent Research Group.

\bibliography{listb}

\end{document}